\documentclass[a4paper,10pt]{article}
\usepackage[utf8]{inputenc}

\title{Generalized Faddeev-Popov Method for a Deformed Supersymmetric  Yang-Mills Theory}
\author{ Paul Weinreb$^1$ and Mir Faizal$^2$\\ 
$^1$Department of Mathematics 
King's College London, \\ 
Strand 
London WC2R 2LS, UK\\ 
$^2$Department of Physics and Astronomy, \\  University of Waterloo,   Waterloo,\\
Ontario N2L 3G1, Canada}

\date{}
\begin{document}

\maketitle

\begin{abstract}
In this paper, we will study the deformation of a three dimensional 
$\mathcal{N} = 2$ supersymmetry gauge theory. 
We will deform this theory by imposing non-anticommutativity. 
This will break the supersymmetry of the theory from $\mathcal{N} = 2$ supersymmetry to 
$\mathcal{N} = 1$ supersymmetry.  
We will address the problem that occurs 
in the Landau gauge due to the existence of multiple solutions to the gauge fixing condition. 
This   will be done by generalizing the Faddeev-Popov method. 
This  formalism is motivated from the Nicolai map in topological field 
theories.  Finally, we will study the extended BRST symmetry 
that occurs  in this theory. 
\end{abstract}
\section{Introduction}
The  studies done on  closed strings in the presence of a constant two form field,  and 
 the gravitational action induced by the bosonic string theory on a
space-filling D-brane with a constant magnetic field,
have motivated a noncommutative deformation of field theories 
 \cite{stri}-\cite{stri1}.  The noncommutative field theory can be constructed 
by replacing all the products of fields in the  action by Moyal products of those fields. This   lead to 
the mixing of ultraviolet and infrared divergences \cite{UV}. These theories are also non-local,
but the non-locality 
is introduced in a controlled manner \cite{NL}. The existence of such noncommutative deformations
have motivated the study of a wider class of deformations for supersymmetric field theories. 
As the coordinate space of supersymmetric theories has Grassmann coordinates, it is possible to impose 
a noncommutative deformation between such coordinates and ordinary spacetime coordinates, along with 
introducing  a non-anticommutative deformation between the Grassmann coordinates 
    \cite{non}-\cite{non1}.  It is possible to deform the   supersymmetric gauge theories 
 using such deformations. 
\cite{ng}-\cite{gn}.  It may be noted that the non-anticommutative deformation of field theories can be 
related to the existence of $R-R$ backgrounds fields \cite{r1}-\cite{1r}.  

In this paper, we will analyse such a deformation for a three dimensional supersymmetric
gauge theory. So, we will promote  the    Grassmann 
coordinates    to   non-anticommutating coordinates. This will break half the supersymmetry of the original theory. 
Thus, if we start from a four dimensional theory with 
$\mathcal{N} = 1$ supersymmetry, this deformation would break the supersymmetry
down to $\mathcal{N} = 1/2$ supersymmetry \cite{12}-\cite{21}. This is because the four dimensional gauge theories 
have enough degrees of freedom to partially break the supersymmetry of the theory. However, if we tried to deform 
a three dimensional theory with $\mathcal{N} = 1$ supersymmetry, we would break all the supersymmetry of the theory. 
So, we will consider a theory with $\mathcal{N} = 2$ supersymmetry in three dimensions, and the non-anticommutativity 
will break half of this supersymmetry. Thus, the 
theory will have $\mathcal{N} = 1$ supersymmetry after this deformation. 
It may be noted that such deformation of supersymmetric field theories has been studied, 
and it has observed that the non-anticommuativity does break the supersymmetry from 
$\mathcal{N} = 2$ supersymmetry to $\mathcal{N} = 1$ supersymmetry 
   \cite{az}-\cite{za}.  
   
 As a gauge  theory has a non-physical degrees of freedom, it cannot be quantized without fixing a gauge. 
The gauge fixing term can be incorporated at a quantum level by adding a ghost term and a gauge fixing term 
to the original classical action. This new action, which is obtained by the adding of the gauge fixing term and the 
ghost term to the original action is invariant under a symmetry called the BRST symmetry. 
\cite{BRST}-\cite{brst1}. It is also invariant under another symmetry which dual to the BRST symmetry called 
the anti-BRST symmetry  \cite{antibrst}.
However,  for 
  non-perturbative gauge-fixing  the   Gribov ambiguity causes a problem \cite{gr}-\cite{gr1}. 
  This is because 
  it is not possible to obtain a unique representative on gauge orbits once large scale field fluctuations  
  in   gauges like the   Landau  gauge. The Gribov ambiguity for supersymmetry theories has been recently studied
  \cite{rg}-\cite{rg1}.

   It may be noted that the Gribov ambiguity is related to the existence of topological properties of 
  field theories \cite{1}.  In fact, it has also been demonstrated that gauge theories can be 
  expressed in terms of    topologically   quantities \cite{2}-\cite{4}.
  This construction has motivated the study of two different kind of 
  topological field theories called the Schwarz type theories \cite{2} and Witten type theories \cite{b}. 
 The Witten type theories   can be related to the existence of  Gribov ambiguity 
  in field theory. This is because  the Witten type theories are related to the 
cohomology, and can have a direct connection with the BRST symmetry in the gauge 
theory. In fact, it has been demonstrated that 
   the Nicolai map is can be constructed in Witten type theories \cite{ni}. The Nicolai map can be used to 
  restricted the path integral to the moduli space of classical solutions \cite{ni}.
  So, the  Nicolai map has also been used to address the Gribov ambiguity 
for usual gauge theories  \cite{GKW}. In this analysis, the BRST symmetry 
of the gauge theory was
extended to include an extended BRST symmetry.  
  We will apply this formalism 
  to three dimensional non-anticommutative Yang-Mills  theory.

The remaining of the paper is organized as
follows. In  section  \ref{a}, we will deform a three dimensional 
supersymmetric Yang-Mills theory theory in $\mathcal{N} =2$ supersaturate formalism. 
This deformation will break the supersymmetry of the theory to $\mathcal{N} =1$ supersymmetry.
In section \ref{b}, we will analyse the quantization of this theory.
We will also address the problem that occurs 
in the Landau gauge due to the existence of multiple solutions to the gauge fixing condition. 
 A formalism that has been motivated by the Nicolai map
in topological field theories will be used to address this problem. In section \ref{c},
we will analyse the extended BRST symmetry for this 
formalism. Finally, in section \ref{d}, we will summarize our results and discuss some possible
extensions of 
this work.  
  
  \section{Non-anticommutativity}\label{a}
We will analyse   a three dimensional   gauge theory in $\mathcal{N} = 2$  
 superspace formalism. This space is parameterized by the coordinates, $(x^\mu, \theta_{1\alpha}, \theta_{2\alpha})$, 
 where   $\theta_{1\alpha} = \theta_{1 +}, \theta_{1-} , $ and $ \theta_{2\alpha} =  \theta_{2 +}, \theta_{2-} $. 
 The generators of $\mathcal{N} = 2$ supersymmetry can   be written as 
 \begin{eqnarray}
   Q_{1a} &=& \partial_{1\alpha} - (\gamma^\mu \theta_1)_\alpha \partial_\mu, 
\nonumber \\
 Q_{2a} &=& \partial_{2\alpha} - (\gamma^\mu \theta_2)_\alpha \partial_\mu.
 \end{eqnarray}
The super-derivatives which commute with these generators of supersymmetry can be written as 
 \begin{eqnarray}
 D_{1\alpha} &=& \partial_{1\alpha} + (\gamma^\mu \theta_1)_\alpha \partial_\mu, 
\nonumber \\
 D_{2\alpha} &=& \partial_{2\alpha} + (\gamma^\mu \theta_2)_\alpha \partial_\mu.
\end{eqnarray}
It is possible to transform the $\theta_1$ and $\theta_2$ to another set of Grassman coordinates 
 \begin{eqnarray}
 \left( \begin{array}{ccc}
\theta_{ a} \\
\bar \theta_{ a} \\
 \end{array} \right) = 
  \left( \begin{array}{ccc}
 x_{11} & x_{12}  \\
 x_{21} & x_{22}  \\
  \end{array} \right) 
 \left( \begin{array}{ccc}
\theta_{1a} \\
\theta_{2a} \\
\end{array} \right),  
 \end{eqnarray}
as long as the $\det (x_{ij}) \neq 0$. Now we  choose $x_{ij}$, such that   
 $ \theta_{a} = (\theta_{1a} + i \theta_{2a})/2, \, \, 
 \bar \theta_{a} = (\theta_{1a} - i \theta_{2a})/2$, and 
  define another set of covariant derivatives as \cite{berm}
\begin{eqnarray}
D_\alpha= \frac{1}{2} ( D_{1\alpha} +  i D_{2\alpha}),&&
 \bar D_\alpha = \frac{1}{2} ( D_{1\alpha} -  i D_{2\alpha}). 
\end{eqnarray}
These covariant derivatives satisfy, 
\begin{eqnarray}
 \{ D_\alpha, \bar D_\beta\} = i(\gamma^\mu \partial_\mu)_{\alpha\beta}, 
&&\{\bar D_\alpha, \bar D_\beta\} =  0,\nonumber \\
   \{ D_\alpha, D_\beta\} = 0. 
&&
\end{eqnarray}
We also define $D^2 = D^\alpha D_\alpha/2$ and $\bar D^2 = \bar D^\alpha \bar D_\alpha/2$. 
This superspace is parameterized by the coordinates, 
 $(x^\mu, \theta^\alpha, \bar \theta^{ {\alpha}})$, where  $\mu = 0, 1,2, 3$, and $\alpha,  {\alpha} = 1, 2$. 
We will impose the 
  non-anticommutative deformation  between $\theta^\alpha$ as 
\begin{equation}
 \{\theta^\alpha, \theta^{\beta} \} = C^{\alpha \beta}. 
\end{equation}
The product of   superfields of $\theta^\alpha$ can be Weyl ordered. This is done by  the ordinary product 
of superfields  by  a star product, which is a  
  fermionic version of the Moyal product 
\begin{equation}
V(x, \theta, \bar \theta) \star   V'(x, \theta, \bar \theta)= V(x, \theta, \bar \theta) 
\exp \left(-\frac{ C^{\alpha\beta}}{2} \overleftarrow{\frac{\partial}{\partial \theta^\alpha }} 
\overrightarrow{\frac{\partial}{\partial \theta^\beta}} \right) V'(x, \theta, \bar \theta), 
\end{equation}
where $V(x, \theta, \bar \theta)$, and $V' (x, \theta, \bar \theta)$ are   supervector fields. 
It may be noted that the Grassmann coordinate $\bar \theta_{ \alpha}$ satisfy, 
\begin{eqnarray}
 \{ \bar \theta_{ {\alpha}}, \bar \theta_{\beta}\} =0, &&  
 \{ \bar \theta_{ {\alpha}},  \theta_\beta\} =0, \nonumber \\
 { [\bar \theta_{ {\alpha}}, x^\mu] }=0,  &&
\end{eqnarray}
and the bosonic coordinates satisfy 
\begin{eqnarray}
 [x^\mu, x^\nu] = \bar \theta \bar \theta C^{\mu \nu} , && [x^\mu, \theta^\alpha ] 
 = i C^{\alpha\beta}\sigma^\mu_{\beta\delta}\bar\theta^{\delta}, 
\end{eqnarray}
where $ C^{\mu\nu} = C^{\alpha\beta}  \epsilon_{\beta \delta}(\sigma^{\mu\nu})_\alpha^\delta $. 
It is possible to write 
\begin{eqnarray}
[  \theta_{\alpha }, y^\mu] =0, &&  
[ \bar \theta_{\alpha}, y^\mu] =0 , \nonumber \\
  {[y^\mu, y^\nu]} =0,  &&
\end{eqnarray}
where $
  y^\mu = x^\mu + i \theta^\alpha  \gamma^\mu_{\alpha\beta} \bar \theta^{\beta},
$
The superfields can be expressed as  functions of $(y^\mu, \theta^\alpha , \bar \theta^{ {\alpha}})$ 
\cite{non}-\cite{non1}. The 
 supervector field $V(y, \theta, \bar \theta)$ in the Wess-Zumino gauge is given by 
\begin{eqnarray}
 V(y, \theta, \bar \theta) &=& - \theta \sigma^\mu \bar \theta A_\mu  + i \theta\theta \bar \theta \bar \lambda
 - i    \bar \theta \bar \theta \theta^\alpha  \left( \lambda_\alpha + \frac{1}{4}\epsilon_{\alpha \beta} 
 C^{\beta \delta} 
 \sigma^\mu_{\delta \rho }
 [\bar\lambda^{\rho}, A_\mu] \right) \nonumber \\ && + \frac{1}{2} 
 \theta\theta \bar \theta \bar \theta  (D - i \partial_\mu A^\mu), 
\end{eqnarray}
where $V^a(y, \theta, \bar \theta)T_a = V(y, \theta, \bar \theta)$. Here $T_a$ are generators of $SU(N)$ Lie algebra, 
\begin{equation}
 [T_a, T_b] = i f_{ab}^cT_c. 
\end{equation}
  We can define the 
  Chiral and anti-Chiral field strength for this theory as 
  \begin{eqnarray}
  W_\alpha &=& - \frac{1}{4} \bar D  \bar D e^{-V}_\star  \star D_\alpha e^V_\star, \nonumber \\ 
  \bar W_{\dot{\alpha}} &=& \frac{1}{4}   D    D e^{-V}_\star  \star \bar D_{\dot{\alpha}} e^V_\star. 
  \end{eqnarray}
The   action  for $\mathcal{N} = 1$ gauge theory can be written as  
\begin{equation}
S_{DSYM} = Tr \int d^3 x d^2 \theta  \, W^\alpha  \star   W_\alpha + Tr \int d^3 x d^2\bar \theta \, 
 \bar W^{\dot{\alpha}} \star   \bar W_{\dot{\alpha}}.   
\end{equation}
It is possible to expand this in component form as 
\begin{eqnarray}
S_{DSYM} &=& Tr \int d^3 x \left[ ( - 4 i \bar \lambda \sigma^\mu D_\mu \lambda - F^{\mu\nu} F_{\mu\nu} + 2 D^2) 
 \right. \nonumber \\  && \left. + Tr \int d^3 x \left( - 2 iC^{\mu\nu} F_{\mu\nu}\bar\lambda\bar \lambda  +
  \frac{C^{\mu\nu}C_{\mu\nu}}{2} (\bar \lambda\bar \lambda)^2 \right) \right]. 
\end{eqnarray}
  \section{Quantization}\label{b}
  In the previous section, we analysed the non-anticommutative deformation of a gauge theory 
  in $\mathcal{N} = 2$ superspace formalism. This deformation broke the supersymmetry of the gauge theory 
  from $\mathcal{N} = 2$ supersymmetry to $\mathcal{N} = 1$ supersymmetry. In this section, we will analyse 
  the quantization of this $\mathcal{N} = 1$ supersymmetry. 
  We will perform this analysis using a covariant formalism. 
We can also express this deformed three dimensional theory using  
the  covariant derivative    \cite{1001} 
  \begin{eqnarray}
 \nabla_A &=& (-i \{\mathcal{D}_\alpha , \bar D_{ {\beta}}\}_\star  , 
\mathcal{D}_\alpha , \bar  D_{ {\alpha}}), 
 \nonumber \\ 
 \exp ( V )_\star \star \nabla_A \star \exp( -V )_\star &=& (-i \{ D_\alpha,\bar  \mathcal{D}_{ 
\beta} \}_\star  , 
  D_\alpha , \bar \mathcal{D}_{ {\alpha}} ), 
\end{eqnarray}
where $ 
 \mathcal{D}_\alpha = \exp (-V)_\star \star   D_\alpha \exp (V )_\star,  $ and $ 
 \bar \mathcal{D}_{ {\alpha}} =
\exp (V)_\star \star   \bar D_{ {\alpha}} \exp(- V )_\star $. 
We can express it using a covarient formalism  as \cite{1001}  
\begin{eqnarray}
 \nabla_A = D_A - i \Gamma_A, 
\end{eqnarray}
where 
\begin{eqnarray}
 D_A &=& ( \partial_{\alpha\beta}, D_\alpha, \bar D_{\beta}), 
\nonumber  \\
 \Gamma_A&=& (\Gamma_{\alpha \beta},\Gamma_\alpha,\bar \Gamma_{ {\alpha}} ). 
\end{eqnarray}
It is possible to express the   Bianchi identity  as
$ [\nabla_{[A}, H_{BC)}\}_\star    =0$, where 
$H_{AB}=  [ \nabla_A, \nabla_B \}_\star   = T^C_{AB}\nabla_C - i F_{AB} $. 
The gauge transformation of this  covariant derivative   are given by 
$\nabla_A \to e^{ i \Lambda}_\star \star \nabla_A \star e^{-i\Lambda}_\star$, 
and $ e^{V}_\star \star \nabla_A \star e^{-V}_\star  
\to e^{i \bar \Lambda}_\star \star e^{V}_\star \star  \nabla_A \star e^{-V}_\star \star e^{ -i \bar\Lambda}_\star$. 
However, it is 
 is possible to construct a covariant derivative in a different  representation, and the 
   gauge transformations   of this covariant derivative is given by 
   \begin{equation}
    \nabla_A \to u \star \nabla_A \star u^{-1}, 
   \end{equation}
where  $u$ is defined as
  $ u = e^{iK}_\star $, and the parameter $K = K^A T_A$ is a real superfield \cite{1001}. 
  The  transformation of the  spinor fields can be written as 
  \begin{eqnarray} 
   \Gamma_\alpha &\to& i u \star \nabla_\alpha \star u^{-1},\nonumber \\
   \bar \Gamma_{ {\alpha}} &\to& i u\star \bar   \nabla_{ {\alpha}}\star u^{-1},\nonumber \\
     \Gamma_{\alpha\beta} &\to& i u \star    \nabla_{ \alpha\beta }  \star u^{-1}. 
  \end{eqnarray}

  As this theory has gauge symmetry, we will have to introduce a ghost term and a gauge 
  fixing term to the original 
 action. However, to calculate non-perturbative effects, we will have to deal with the Gribov ambiguity. So, 
 we will   apply the extended BRST symmetry to this theory \cite{GKW}. 
 Thus, for a gauge fixing condition $F[^g \Gamma] =0, $ and $ \bar F[^g \bar \Gamma] =0 $,
 where $g$ is an element of $SU(N)$. For the Landau gauge, we can write 
 \begin{eqnarray}
  F[\Gamma] = D^a \Gamma_a =0, &&   \bar F[\bar \Gamma] = \bar D^a \bar \Gamma_a =0.
 \end{eqnarray}
The Faddeev-Popov operator can   be written as 
\begin{eqnarray}
 M_F[\Gamma] = \left(\frac{\delta F[^g \Gamma]}{\delta g}\right), && 
 \bar M_F[\bar\Gamma] = \left(\frac{\delta \bar F[^g\bar  \Gamma]}{\delta g}\right).
\end{eqnarray}
In the  standard Faddeev-Popov method,  the following expression is used,  
\begin{eqnarray}
 1&=&\int \mathcal{D}_g\Delta_F[^g\Gamma]\delta[F[^g\Gamma]] + \int \mathcal{D}_g\bar \Delta_F[^g\bar
 \Gamma]\delta[\bar F[^g\bar\Gamma]],  
 \end{eqnarray}
However, as we can have multiple solutions in the Landau gauge, we can generalize the 
 standard Faddeev-Popov to 
multiple solutions as follows \cite{GKW},
\begin{eqnarray}
 N_F[\Gamma,  \bar \Gamma]&=&\int \mathcal{D}_g\Delta_F[^g\Gamma]\delta[F[^g\Gamma]] + 
 \int \mathcal{D}_g\bar \Delta_F[^g\bar \Gamma]\delta[\bar F[^g\bar \Gamma]] 
 \end{eqnarray}
 where $ N_F[\Gamma,\bar \Gamma]$  denotes  the number solutions   corresponding to 
 a gauge fixing condition.

 It is known that for usual gauge theories, the fundamental modular region is a 
 unique representation of every gauge orbit \cite{x}-\cite{y}. Even though the theory considered 
 here is a supersymmetric Yang-Mills theory, it is a gauge theory and hence we expect such 
 a unique representation of every supergauge orbit. Furthermore, it is possible to generalize the 
 arguments used for obtaining these results for a supersymmetric theory. 
It may be noted that even though this theory has $\mathcal{N} = 1$ supersymmetry, 
 the gauge symmetry for this theory is different from both 
 the un-deformed  Yang-Mills theory with  $\mathcal{N} = 1$ supersymmetry and 
 the un-deformed  Yang-Mills theory with  $\mathcal{N} = 2$ supersymmetry. This is because 
 the gauge transformations are defined in terms of the deformed superspace coordinates, and 
 thus involve fermionic version of the Moyal product. However, as this theory has a gauge symmetry, 
 we can in principle define  unique representation of every supergauge orbit on deformed superspace. 
So,   we will   
 denote the 
  fundamental modular region  $\Lambda$, as  the set of absolute minima of the functional
\begin{eqnarray}
 V_{\Gamma, \bar \Gamma} =\int d^3x d ^2 \theta (^g\Gamma)^2_\star  + \int d^3x d ^2 \bar \theta (^g\bar \Gamma)^2_\star.  
 \end{eqnarray}
 The stationary points of $V_{\Gamma, \bar \Gamma}$ satisfy the Landau gauge condition, and the 
 boundary of the  fundamental modular region  $\partial \Lambda$ is the set of degenerate absolute minimum 
 of $V_{\Gamma, \bar \Gamma}$. The  fundamental modular region lies within the Gribov region, and  the 
 operators $M_F [\Gamma]$ and $\bar M_F [\bar \Gamma]$ obtain zero modes in the boundary of the Gribov region, 
 which is called the Gribov horizon. The gauge orbits  can be labeled using this 
 fundamental modular region, and denoted  $\Gamma (v), \bar \Gamma(v)$ as configurations in the 
 fundamental modular region. So, we have $N_F[\Gamma, \bar \Gamma]$, as
 every orbit crosses the fundamental modular region only once. 
 Since the integrand is positive the minima of $V_{\Gamma, \bar \Gamma}$ 
 are those $\Gamma_\alpha, \bar \Gamma_\alpha$  
 satisfying the Landau gauge condition $D^\alpha\Gamma^\alpha=0$ and  $\bar D^\alpha\bar \Gamma^\alpha=0$. 
 The boundary $\partial \Lambda$ is the set of degenerate absolute 
 minima of  $V_{\Gamma, \bar \Gamma}$.   
 
The expectation value of a gauge invariant operator $\mathcal{O}_\star$ defined on the deformed superspace, 
can be  written as 
\begin{eqnarray}
\langle\mathcal{O}{[\Gamma, \bar \Gamma]_\star}\rangle
&=&\frac{\int \mathcal{D}  \mathcal{O}
[\Gamma,\bar\Gamma]_\star e^{-S_{DSYM}}}{\int \mathcal{D}  e^{-S_{DSYM}}}, 
\end{eqnarray}
where $\mathcal{D} =\mathcal{D}[ \Gamma (v)   \bar \Gamma (v)] $ is a suitably defined 
  measure for the path integral. 
This is well defined if there are unique solutions to the gauge fixing condition. 
Since $N_F $ is finite, we obtain we write
\begin{eqnarray}
\langle\mathcal{O}{[\Gamma, \Gamma]_\star}\rangle&=&
\left[\int \mathcal{D}    {N_F[\Gamma, \bar \Gamma]}^{-1}
\int \mathcal{D} g \delta(F[^g\Gamma])|\det M_F[^g\Gamma]\right.\nonumber \\  &&\left.
\delta(\bar F[^g\bar\Gamma])|\det \bar M_F[^g\bar\Gamma]|
\mathcal{O}[\Gamma, \bar \Gamma]_\star e^{-S_{DSYM[\Gamma, \bar \Gamma]}
[\Gamma, \bar \Gamma]}\right]_\star\nonumber \\ && \times   \left[\int \mathcal{D}   {N_F[\Gamma, \bar \Gamma]}^{-1}
\int \mathcal{D}g\delta(F[^g\Gamma])|\det M_F[^g\Gamma]
\right.\nonumber \\  && \left. \delta(\bar F[^g\bar \Gamma])|\det \bar M_F[^g\bar\Gamma]
|e^{-S_{DSYM}[\Gamma, \bar \Gamma]}\right]_\star^{-1}.
\end{eqnarray}
So,  we have  
$N_F[\Gamma (v), \bar \Gamma (v)] = N_F[^g\Gamma (v), ^g \bar \Gamma (v)]=N_F[\Gamma, \bar \Gamma] $. 
So, we can write   
\begin{eqnarray}
\langle\mathcal{O}{[\Gamma, \bar \Gamma]_\star}\rangle&=&
\left[\int \mathcal{D}  \delta(F[\Gamma])|\det M_F[\Gamma] |\delta(\bar F[\bar\Gamma])\right.\nonumber \\
&& \left.|\det \bar M_F[\bar\Gamma]|\mathcal{O}[\Gamma, \bar \Gamma]
e^{-S_{DSYM}[\Gamma, \bar \Gamma]} \right]_\star \nonumber \\ && \times \left[\int \mathcal{D}  \delta(F[\Gamma])
|\det M_F[\Gamma]|\delta(\bar F[\bar\Gamma])\right.\nonumber \\  && \left.
|\det \bar M_F[\bar\Gamma]|e^{-S_{DSYM}[\Gamma, \bar \Gamma]}\right]_\star^{-1}.
\end{eqnarray}
It may be noted that this expression involves the fermionic version of the Moyal product.
This expression can be written in terms of the power  series involving the deformation matrix
$C^{\alpha \beta}$. The first term in the series will correspond to the usual un-deformed supersymmetric 
Yang-Mills theory. Thus, the expectation value of a gauge invariant operator also receives corrections 
from the deformation of the Yang-Mills theory. Furthermore, in absence of supersymmetry, this expression 
reduces to the expression for the usual Yang-Mills case. This can be seen by setting by making all the fermionic 
fields in the expression to vanish. However, this will be different from breaking all the supersymmetry by 
imposing two non-anticommutative deformations. The expectation value of gauge invariant operators in this case, 
will be different from the expectation value of gauge invariant operators of the usual 
Yang-Mills theory.

\section{Symmetries}\label{c}
In the previous section, we analysed the quantization of a three dimensional non-anticommutative gauge theory.
We generalized the 
 standard Faddeev-Popov method  to incorporate the existence of 
multiple solutions to the gauge fixing condition. This was done to   address the existence of   
non-perturbative effects.   In this section, we will analyse the BRST symmetry for this generalized 
Faddeev-Popov method.  
The partition function for this generalized Faddeev-Popov method,  can be written as 
\begin{eqnarray}
Z_{gf}&=&\left[\int \mathcal{D}  \delta(F[\Gamma])|\det M_F[\Gamma] |\delta(\bar F[\bar\Gamma])\right.\nonumber \\
&& \left.|\det \bar M_F[\bar\Gamma]|
e^{-S_{DSYM}[\Gamma, \bar \Gamma]} \right]_\star.
\end{eqnarray}
This is valid for non-perturbative field theory, as it takes the modulus of the 
determinant into account. Thus, we can write \cite{GKW}
\begin{eqnarray}
 |\det M_F [\Gamma] |_\star = [\rm{sgn}(\det M_F [\Gamma]) \det M_F [\Gamma]]_\star,
 \nonumber \\  |\det \bar M_F [\bar \Gamma]_\star = [\rm{sgn}(\det \bar M_F [\bar \Gamma]) \det \bar M_F [\bar\Gamma]]_\star.  
\end{eqnarray}
We can express the action corresponding to $ \det M_F [\Gamma] $ and $\det \bar   M_F [\bar\Gamma]$ as follows, 
\begin{eqnarray}
S_{det}&=&\int d^3 x d^2 \theta \left[-b^a \star D^\alpha\Gamma^a_\alpha 
+\frac{\xi}{2}b^a \star b^a+  \tilde c^a \star M_F^{ab} \star c^b\right]
\nonumber \\&& + \int d^3 x d^2\bar  \theta \left[
-\bar b^a \star \bar D^\alpha  \bar \Gamma^a_\alpha +\frac{\xi}{2} \bar b^a \star
\bar b^a+ \bar{\tilde c}^a \star \bar M_F^{ab}\star  c^b\right]. 
\end{eqnarray}
Here the ghosts and anti-ghosts are denoted by $c^a, \bar c^a$ and $\tilde c^a,  \bar{\tilde c}^a$, respectively. 
The Nakanishi-Lautrup  auxiliary fields are denoted by $b^a, \bar b^a$.  
So,  we obtain 
\begin{eqnarray}
\lim_{\xi\rightarrow0}\int \mathcal{D}   e^{-S_{det}}&=&[\delta (F[\Gamma]) \det M_F [\Gamma] 
\delta (\bar F[\bar \Gamma]) \det \bar M_F [\bar \Gamma] ]_\star
\end{eqnarray}
where the measure $\mathcal{D}$ includes an integral over the ghosts, anti-ghosts and auxiliary fields. 
We can also write the 
 action  corresponding to  $\rm{sgn}(\det   M_F [\Gamma])$ and $\rm{sgn}(\det \bar M_F [\Gamma])$ as
 \begin{eqnarray}
 S_{sgn}  &=&\int d^3 x d^2 \theta \left[ 
 i B^a \star M_F^{ab}\star \phi^b-i\tilde d^a \star M_F^{ab}\star d^b+ \frac{1}{2}B^a\star B^b\right] 
 \nonumber \\
 && +\int d^3 x d^2 \bar \theta \left[ 
 i \bar B^a \star \bar   M_F^{ab}\star \bar \phi^b
 -i\bar{\tilde d}^a \star\bar   M_F^{ab}\star \bar d^b + \frac{1}{2}\bar B^a\star \bar B^b \right] .
 \end{eqnarray}
 
Here $ d^a, \bar d^a, , \tilde d^a, \bar {\tilde d}^a$ 
are new Grassman odd  superfields and 
$\phi^a, \bar \phi^a$, $B^a,\bar B^a$ are new auxiliary Grassman even superfields.
Completing the square, the $B$ field can be integrated out, and so we can define the effective action as
\begin{eqnarray}
S'_{sgn}&=& \int d^3 x d^2 \theta \left[\frac{1}{2}\phi^a\star ((M_F)^T)^{ab}\star M_F^{bc}\star 
\phi^c-i\tilde d^a\star M_F^{ab}\star d^b\right]
\nonumber \\ &&
\int d^3 x d^2 \bar\theta \left[\frac{1}{2}\bar \phi^a\star ((\bar M_F)^T)^{ab}\star \bar M_F^{bc}\star 
\bar \phi^c-i\bar{\tilde{d}}^a\star\bar M_F^{ab}\star \bar d^b\right].
\end{eqnarray} 
Thus, the partition function can be written as 
\begin{eqnarray}
\nonumber
\mathcal{Z}_{gf}=\int \mathcal{D}  N_F[\Gamma, \bar \Gamma]^{-1} e^{-S_{DSYM}-S_{det}-S_{sgn}}. 
\end{eqnarray} 

It may be noted that all these superfields are defined on deformed superspace. So, just like the gauge transformations, 
the BRST and the anti-BRST transformations of these superfields will also involve 
the fermionic version of the Moyal product. However, apart from the difference the expression in the superspace take 
similar forms. In fact, these deformed expression can be expressed as of power series involving 
the deformation matrix $C^{\alpha \beta}$. The first term in this series will correspond to the usual 
BRST and the usual anti-BRST transformations. So, even though they appear to have similar form in the superspace, 
the component form of the  BRST and the anti-BRST transformations will
  look very   different for the deformed and 
the un-deformed Yang-Mills theory. Furthermore, the   BRST and the anti-BRST transformation for the usual Yang-Mills 
theory can be obtained by making all the fermionic field to vanish in the supersymmetric Yang-Mills theory. 
However, it is also possible to obtain a Yang-Mills theory without supersymmetry by breaking all the supersymmetry 
by imposing two non-anticommutative deformations. In this case, the BRST and the anti-BRST transformation will be 
different from the BRST and the anti-BRST transformations for the usual Yang-Mills theory. 

The standard BRST   transformations   for the gauge fields 
$s$ and $\bar s$ can be written as  $s \Gamma_\alpha^a = \nabla_\alpha^{ab} \star  c^b$ and 
$s \bar \Gamma_\alpha^a = \bar \nabla_\alpha^{ab} \star \bar c^b$. 
We can also write the anti-BRST transformations for the gauge fields as 
$\tilde s \Gamma_\alpha^a = \nabla_\alpha^{ab} \star  \tilde c^b$ and 
$s \bar \Gamma_\alpha^a = \bar \nabla_\alpha^{ab} \star \bar{\tilde c}^b$. 
The BRST transformation of the ghosts is given by 
$s c^a = - f_{bc}^a c^b \star c^c/2$ and $s \bar c^a = - f_{bc}^a \bar c^b \star \bar c^c/2$.
The anti-BRST transformation of anti-ghosts is given by 
$\tilde s \tilde c^a = - f_{bc}^a \tilde  c^b \star \tilde  c^c/2$ and $\tilde  s \bar {\tilde c}^a =
- f_{bc}^a \bar {\tilde c}^b \star \bar {\tilde c}^c/2$. The BRST transformation of anti-ghosts is given by 
$s c^a = b^a $ and $s \bar c^a = \bar b^a$.
The anti-BRST transformation of ghosts is given by $s \tilde c^a = \tilde b^a $ and $s \bar {\tilde{c}}^a 
= -\bar b^a$. 
Apart from this the BRST and anti-BRST transformation of all the other auxiliary fields vanishes, 
$s b^a = 0$, $ s \bar b^a =0$ and $\tilde s b^a = 0$, $ \tilde s \bar b^a =0$. 
Apart from these BRST and anti-BRST transformations, this action is also invariant under a double BRST 
and an anti-BRST transformations. The double BRST transformations are given by 
$t \phi^a = d^a$, $t \bar \phi^a = \bar d^a$ and $t \tilde d^a = B^a$, $t \bar  {\tilde d}^a = \bar B^a$. 
The double BRST transformation of all the other fields vanishes, $t d^a =0$, $t\bar d^a = 0$ and 
$t B^a =0$, $ t \bar B^a =0$. The double anti-BRST transformations are given by 
$\tilde t \phi^a = \tilde d^a$, $\tilde  t \bar {\phi}^a = \bar {\tilde d}^a$ and $\tilde  
t   d^a = - B^a$, $\tilde  t \bar  {  d}^a =- \bar B^a$. 
The double BRST transformation of the all the other fields vanishes, $\tilde  t \tilde 
d^a =0$, $\tilde  t\bar {\tilde d}^a = 0$ and 
$\tilde  t B^a =0$, $ \tilde  t \bar B^a =0$.   
So,  we can define fields $\Gamma^a_i = (\Gamma^a_\alpha, \phi^a),
\bar \Gamma^a_i = (\bar \Gamma^a_\alpha, \bar\phi^a), \mathcal{C}^a_i = (c^a, d^a), \bar \mathcal{C}^a_i = 
(\bar c^a, \bar d^a), \tilde \mathcal{C}^a_i = (\tilde c^a, \tilde d^a), \bar {\tilde \mathcal{C}}^a_i = 
(\bar {\tilde{c}}^a, \bar {\tilde{d}}^a), \mathcal{B}^a_i = (b^a, B^a), \bar \mathcal{B}^a_i = (\bar b^a,\bar B^a)$. 
The BRST transformations can   be written as $\mathcal{S} \Gamma^a_i = (\nabla^{ab} \star \mathcal{C}^b)_i, 
\mathcal{S} \bar \Gamma^a_i = (\bar \nabla^{ab} \star \bar \mathcal{C}^b)_i,  
\mathcal{S} \mathcal{C}_i^a = Y^{jk}_i  f^a_{bc} \mathcal{C}^b_j  \star   \mathcal{C}^c_k, 
\mathcal{S} \bar \mathcal{C}_i^a = Y^{jk}_i  f^a_{bc} \bar \mathcal{C}^b_j  \star  \bar  \mathcal{C}^c_k, 
\mathcal{S}\tilde  \mathcal{C}^a = \mathcal{B}^a_i, \mathcal{S}\bar {\tilde \mathcal{C}}^a_i = \mathcal{B}^a_i, 
\mathcal{S} \mathcal{B}^a_i =0, \mathcal{S} \bar \mathcal{B}^a_i =0,  
$ where $Y^1_{11} = 1, Y^i_{jk} =0, $ if $i, j, k \neq 1$. It is possible to write  \cite{GKW}
\begin{eqnarray}
 S_{det} + S_{sgn} = Tr \int d^3 x d^2 \theta   \mathcal{S} \mathcal{U}_\star  + 
 Tr \int d^3 x d^2 \bar \theta   \mathcal{S} \bar \mathcal{U}_\star  , 
\end{eqnarray}
where  $\mathcal{U}_\star  = \rm{diag}(\tilde c^a \star F^a, \tilde d^a \star (i M_F^{ab} \star \phi^b +  B^a/2  )   ) $ and 
$\bar \mathcal{U}_\star  = \rm{diag}(\bar{\tilde c}^a \star \bar F^a, \bar {\tilde d}^a \star 
(i \bar M_F^{ab} \star  \bar \phi^b +  \bar B^a/2)   ) $. The anti-BRST transformation can   be written as 
 $\tilde \mathcal{S} \Gamma^a_i = (\nabla^{ab} \star \tilde \mathcal{C}^b)_i, 
\tilde \mathcal{S} \bar \Gamma^a_i = (\bar \nabla^{ab} \star \bar {\tilde\mathcal{C}}^b)_i,  
\tilde \mathcal{S} \tilde \mathcal{C}_i^a = Y^{jk}_i  f^a_{bc} \tilde \mathcal{C}^b_j  \star   \tilde \mathcal{C}^c_k, 
\tilde \mathcal{S} \bar {\tilde\mathcal{C}}_i^a = Y^{jk}_i  f^a_{bc} \bar{\tilde \mathcal{C}}^b_j  \star  \bar  {\tilde\mathcal{C}}^c_k, 
\tilde \mathcal{S}\tilde  \mathcal{C}^a_i = -\mathcal{B}^a_i, \tilde \mathcal{S}\bar { \mathcal{C}}^a_i = -\mathcal{B}^a_i, 
\tilde \mathcal{S} \mathcal{B}^a_i =0, \tilde \mathcal{S} \bar \mathcal{B}^a_i =0. $ 
We can write \cite{GKW}
\begin{eqnarray}
 S_{det} + S_{sgn} = Tr \int d^3 x d^2 \theta   \mathcal{S} \tilde \mathcal{S}\mathcal{W}_\star  + 
 Tr \int d^3 x d^2 \bar \theta   \mathcal{S}\tilde \mathcal{S} \bar \mathcal{W}_\star  , 
\end{eqnarray}
where $\mathcal{W}_\star  =\rm{diag}( \Gamma^{\alpha a} \star \Gamma^a_\alpha, \phi^a \star M_F^{ab} \star  \phi^b, 
\tilde d^a \star d^a)$
and $\bar \mathcal{W}_\star  =\rm{diag}( \bar\Gamma^{\alpha a} \star \bar \Gamma^a_\alpha,
\bar\phi^a \star \bar M_F^{ab} \star  \bar \phi^b, 
\bar {\tilde{d}}^a \star \bar d^a)$. Thus, we are able to formulate the modulus of the determinant in Landau gauge 
in terms of a Lagrangian. However, this procedure   also holds for non-perturbative phenomena. 

\section{Conclusion}\label{d}
In this paper, we analysed a three dimensional supersymmetric Yang-Mills theory on deformed superspace. We deformed 
the superspace by imposing non-anticommutativity. The original theory has $\mathcal{N} =2 $ supersymmetry. The 
deformation of this theory broke half of the supersymmetry of the original theory. Thus, the final  theory had only 
$\mathcal{N} = 1$ supersymmetry.   We  analysed the quantization of this theory, and 
  addressed the problem that occurs
 due to the Gribov ambiguity. This was done by 
  generalizing the 
 standard Faddeev-Popov method. This generalized Faddeev-Popov method was used for analysing 
 the existence of 
multiple solutions to the gauge fixing condition. Thus, non-perturbative effects could 
be   address using this generalized Faddeev-Popov method.  We derived an expression for 
calculating the expectation values of  gauge invariant operators. A partition function for this deformed 
theory was also constructed. We 
analysed the BRST and the anti-BRST symmetries 
this partition function. It was demonstrated that apart from being invariant under the usual BRST and 
the usual anti-BRST transformations, this partition function also invariant under double
BRST and double anti-BRST 
transformations.  We were able to combine  the usual BRST and anti-BRST transformation, with these new
BRST and anti-BRST transformations. These new BRST and anti-BRST transformations, 
take into account the existence of multiple solutions to the gauge fixing conditions, and 
so the   results of this paper can be used for analysing 
different aspects of non-perturbative phenomena.

It may be noted that a similar analysis can be done for theories with 
higher amount of supersymmetry. It would be interesting to apply this formalism for studying supersymmetric 
theories with  boundaries. This is because it is possible to construct a three dimensional theory with 
$\mathcal{N} = 1/2$ supersymmetry by combining the boundary effects with non-anticommuativity \cite{za}. 
However, the quantization of this theory has not been studied. It would be interesting to analyse the 
effect of boundaries on the results obtained in this paper. 
It may be noted that the existence of   Gribov ambiguity  is related to the gauge fixing procedure for
quantizing Yang-Mills theories, and this problem is usually addressed in the  Gribov-Zwanziger formalism 
\cite{z1}-\cite{z}. 
It is possible  to analyse effects coming from  the existence of the Gribov
copies in a local  way using this formalism. 
This formalism has been used to analyse the  infrared behavior of the gluon and ghost propagator. 
In fact, the Gribov-Zwanziger formalism has been used for studding  the zero momentum 
value of the gluon propagator. These propagators have been used for analysing the 
 the spectrum of gauge theories \cite{z2}-\cite{2z}.  The supersymmetric generalization of the 
 Gribov-Zwanziger formalism  has also been performed \cite{supe}. 
 This has been used for analysing 
 existence of the   condensate and  vanishing of the vacuum energy. 
The  renormalization of supersymmetric Yang-Mills theory with $\mathcal{N}=1$ supersymmetry 
 was  analysed using the Gribov-Zwanziger formalism \cite{super}. This was done by  using the Landau condition. 
 The proof of renormalizability of this theory to all orders was studied using an
 algebraic renormalization procedure. It was demonstrated that only  three renormalization constants are needed for 
 this theory.  
  In fact, the   non-renormalization theorem  in the Landau gauge were analysed using the 
  Gribov-Zwanziger formalism. The renormalization  factors for a  non-linear
realization of the supersymmetry were also studied in this formalism. 
 It will be interesting to analyse the effect of non-anticommutative deformation on these results. This  
 can be done by analysing a deformed supersymmetric Yang-Mills theory in  Gribov-Zwanziger formalism.
 Thus, we can study a non-anticommutative a four dimensions supersymmetric Yang-Mills theory 
 in $\mathcal{N}=1$ superspace formalism. The non-anticommutativity will break the supersymmetry of the theory 
 from $\mathcal{N}=1$ supersymmetry  to $\mathcal{N}=1/2$ supersymmetry. 
 It will also be interesting to perform a similar analysis for a deformed three dimensional theory in 
 $\mathcal{N}=2$ superspace formalism. Here the non-anticommuativity will break the supersymmetry 
 of the theory from $\mathcal{N}=2$ supersymmetry  to $\mathcal{N}=1$ supersymmetry. 
 It will be interesting to analyse 
 the effect of this supersymmetry breaking on the existence of the   condensate and  vanishing of the vacuum energy.

\end{document}